\documentclass[
preprint,
tighten,
amsmath,
amssymb,
aps,
twocolumn,
]{aastex7}

\usepackage{graphicx}
\usepackage{dcolumn}
\usepackage{bm}
\usepackage{color}
\usepackage{natbib}
\usepackage{comment}
\usepackage{verbatim}

\newcommand{\msig}{$M_{\mathrm{BH}}-\sigma_\star$}
\newcommand{\mlum}{$M_{\mathrm{BH}}-L_{\mathrm{bul}}$}
\newcommand{\mmass}{$M_{\mathrm{BH}}-M_{\mathrm{bul}}$}
\newcommand{\mmtot}{$M_{\mathrm{BH}}-M_\star$}

\newcommand{\mbh}{$M_{\mathrm{BH}}$}

\newcommand{\kms}{km s$^{-1}$}

\begin{document}

\shortauthors{Cohn et al.}

\title{Evidence for evolutionary pathway-dependent black hole scaling relations}

\author[0000-0003-1420-6037]{Jonathan H. Cohn}
\affiliation{Department of Physics and Astronomy, Dartmouth College, 6127 Wilder Laboratory, Hanover, NH 03755, USA}
\email{jonathan.cohn@dartmouth.edu}

\author[0009-0004-9516-9593]{Emmanuel Durodola}
\affiliation{Department of Physics and Astronomy, Dartmouth College, 6127 Wilder Laboratory, Hanover, NH 03755, USA}
\email{emmanuel.a.durodola.gr@dartmouth.edu}

\author[0009-0005-9002-4800]{Quinn O. Casey}
\affiliation{Department of Physics and Astronomy, Dartmouth College, 6127 Wilder Laboratory, Hanover, NH 03755, USA}
\email{quinn.o.casey.gr@dartmouth.edu}

\author[0000-0003-3216-7190]{Erini Lambrides}
\affiliation{NASA-Goddard Space Flight Center, Code 662, Greenbelt, MD 20771, USA}
\email{erini.lambrides@nasa.gov}

\author[0000-0003-1468-9526]{Ryan C. Hickox}
\affiliation{Department of Physics and Astronomy, Dartmouth College, 6127 Wilder Laboratory, Hanover, NH 03755, USA}
\email{ryan.c.hickox@dartmouth.edu}

\correspondingauthor{Jonathan H. Cohn}
\email{jonathan.cohn@dartmouth.edu}

\begin{abstract}

Recent observations have identified an abundance of high-redshift active galactic nuclei (AGN) with supermassive black holes (BHs) that are over-massive compared to the local BH mass$-$total stellar mass (\mmtot) relation.
\mbh\ measurements at high-$z$ are critical for probing the growth histories of BHs and their host galaxies, including BH seeding and evolution of the \mmtot\ relation.
However, BH masses in high-$z$ AGN are generally estimated from single-epoch measurements, which are anchored to local relations based on reverberation mapping and carry large systematic uncertainties.
Alternate \mbh\ detection methods such as dynamical measurements are more reliable but currently only possible in the local Universe or with strongly lensed systems.
Recently, dynamical \mbh\ measurements were made in a $z\sim2$ lensed quiescent galaxy as well as a sample of six local galaxies identified as likely relics of common quiescent red nugget galaxies at cosmic noon.
We compare the $z\sim2$ red nugget and relic BHs to recent results for $4<z<11$ AGN, quasars, and Little Red Dots.
Intriguingly, the $z\sim2$ galaxy and local relic galaxies all lie on both the local \mmtot\ relation for bulges and the $4<z<7$ \mmtot\ relation.
Our results suggest the \mmtot\ relation for bulges was likely in place at high-$z$ and indicate careful consideration of different evolutionary pathways is needed when building BH scaling relations.
While improvements to \mbh\ estimates in AGN will increase our confidence in high-$z$ BH masses, detecting BHs in relic galaxies and lensed galaxies presents a complementary probe of the high-$z$ relations.
\end{abstract}

\section{Introduction\label{intro}}

Supermassive black holes (BHs) are important galactic building blocks, existing at the centers of essentially all massive galaxies in the local Universe (e.g., \citealt{Saglia2016}).
Dynamical measurements of BH masses (\mbh) have revealed they correlate with large-scale galactic properties, extending well outside the BHs' gravitational spheres of influence (SOI) and implying BHs and host galaxies must co-evolve (e.g., \citealt{Kormendy2013}).
Correlations with \mbh\ have been found with galaxies' total mass ($M_\star$), bulge mass ($M_\mathrm{bul}$), bulge luminosity ($L_\mathrm{bul}$), and stellar velocity dispersion ($\sigma_\star$), among other properties (e.g., \citealt{Kormendy2013,ReinesVolonteri2015,Saglia2016}).
Studying BHs in the early Universe is critical for developing our understanding of co-evolution and BH seeding mechanisms (at $z\gtrsim10$; e.g., \citealt{Volonteri2009}).
However, due to sensitivity and resolution requirements, dynamical BH detections are only possible at distances $\lesssim$150 Mpc or in rare gravitationally lensed systems at high-$z$, making it difficult to explore the evolution of the scaling relations.

Fortunately, there are alternate methods of estimating BH masses, based on active galactic nuclei (AGN).
The most reliable estimates come from reverberation mapping (RM), which measures the time delay in the arrival of flux variations in the AGN broad-line region (BLR) after observing variations in the continuum (e.g., \citealt{Shen2024}).
RM measurements have revealed a correlation between the BLR size and AGN luminosity ($R_\mathrm{BLR}-L$), allowing for estimates of \mbh\ from single observations of an AGN's line luminosity and width, known as single-epoch measurements (e.g., \citealt{Peterson2004}).
These single-epoch measurements enable \mbh\ estimates at very high redshifts, enabling the first studies of BH scaling relation evolution (e.g., \citealt{Izumi2019,Pensabene2020,Larson2023,Bogdan2024,Maiolino2024}).

Recent JWST observations have revealed a population of optically red, compact, high redshift ($5\leq z\leq10$) galaxies.
These X-ray quiet galaxies \citep{Ananna2024}, termed ``Little Red Dots" (LRDs), appear to be ubiquitous in the high redshift Universe.
Spectroscopic analysis of LRDs has revealed a subset population with broad ($\sim$ 4000 km s$^{-1}$) H$\alpha$ emission \citep{Greene2024}, suggesting AGN activity.
Using single-epoch measurements, several studies \citep{Greene2024, Rinaldi2024, Kokorev2024, Killi2024, Kocevski2024, Maiolino2024, Labbe2025} have estimated BH masses hosted by LRDs to range between 10$^6$ and 10$^9$ M$_\odot$.
These \mbh\ estimates are significantly overmassive compared to the local population, with BHs in LRDs 2 to 4 orders of magnitude more massive compared to local galaxies of similar stellar masses \citep{Pacucci2023,Durodola2025}.
These results have revealed a previously unknown tension between the high-$z$ BH population and the local Universe, which astronomers are now working diligently to understand.

However, several issues make it difficult to draw conclusions about the evolution of the BH$-$host galaxy relations when only using single-epoch measurements.
There is an intrinsic factor of $\sim$2 systematic uncertainty in RM measurements due to the geometric virial factor $f$ \citep{Shen2024}.
Furthermore, there are questions surrounding whether the $R_\mathrm{BLR}-L$ relations observed locally change with redshift, increasing uncertainty in single-epoch measurements derived from these relations at high-$z$.
Exacerbating matters, there has been disagreement in the literature regarding whether high-$z$ samples appear to host over-massive BHs due to biases in the sample selection, as brighter AGN with more massive BHs are easier to detect (e.g., \citealt{Li2024}).
Extricating accurate stellar masses from systems with AGN or quasars (QSOs) is also difficult due to AGN contamination.
This may result in biased stellar masses, especially in fainter, AGN-dominated systems \citep{Zhuang2024}.
All these effects compound to cast doubt on the placement of high-$z$ single-epoch BHs on the \mmtot\ relation.
Complementary methods of probing BHs in the early Universe are required to clarify whether high-$z$ systems truly house over-massive BHs.

Recently, a sample of local compact galaxies were identified to have not undergone any \textit{in-situ} stellar growth or major mergers since $z\sim2$, making them likely relics representative of common massive, quiescent red nugget galaxies at cosmic noon \citep{Yildirim2017}.
Red nuggets at $z\sim2$ usually evolve to form the cores of local early-type galaxies (ETGs), growing primarily through dry mergers that increase stellar mass in the galaxy outskirts \citep{Dokkum2010}.
Such local massive ETGs house the most massive BHs in the nearby Universe \citep{Kormendy2013}, and studying the BHs in their progenitors could provide direct insights into coevolution.
Local relics of red nugget galaxies thus provide an ideal window into this important population of $z\sim2$ BHs.
Additionally, as these quiescent relics exist in the local Universe, their BHs can be studied with robust dynamical \mbh\ measurements, insensitive to the uncertainties affecting \mbh\ estimates in AGN.
Dynamical \mbh\ measurements have been made in six red nugget relics to date, revealing a population of BHs that appears to be over-massive compared to the \mmass\ and \mlum\ relations \citep{Walsh2015,Walsh2016,Walsh2017,Cohn2021,Cohn2023,Cohn2024}.
Enticingly, a new stellar-dynamical \mbh\ measurement in a gravitationally lensed quiescent galaxy at $z=1.95$ has also found that it is overmassive on \mmass\ \citep{Newman2025}.
This could indicate that before $z\sim2$, the cores of ETGs had largely finished growing, and between $0<z<2$, dry mergers in ETGs increased their stellar mass without driving additional BH growth, bringing them onto the local BH scaling relations \citep{Mateu2015}.
However, the red nugget sample has yet to be directly compared to high-$z$ AGN \mbh\ measurements or the \mmtot\ relations (locally or at high redshift).

In this letter, we investigate how the red nuggets at cosmic noon may be used as a stepping stone to help clarify whether BHs at high redshift are overmassive compared to the local relations.
To do this, we compare the relic galaxies to a selection of broad-line AGN, luminous QSOs, and LRDs at $4<z<11$, as well as the local and $4<z<7$ \mmtot\ relations from \citet{ReinesVolonteri2015} and \citet{Pacucci2023}, respectively.
We detail our sample selection in \S\ref{sample}, compare the BH masses in our sample to local and high-redshift scaling relations in \S\ref{results}, and discuss the implications of our results for our understanding of BH$-$host galaxy coevolution in \S\ref{conclusions}.

\section{\label{sample}Sample selection}

We compare several datasets representing \mbh\ measurements from $z\sim2$ to the early Universe.
To ensure our analysis of high-$z$ BH masses is not solely reliant on single-epoch measurements, we first select a sample of six red nugget relic galaxies with dynamical \mbh\ measurements.
\citet{Cohn2021,Cohn2023,Cohn2024} performed molecular gas-dynamical \mbh\ modeling with ALMA CO(2$-$1) band 6 observations at resolutions of $\sim$1$-$2$\times$ the BH SOI in three of these galaxies, and \citet{Walsh2015,Walsh2016,Walsh2017} performed stellar-dynamical \mbh\ modeling of high resolution Gemini North/NIFS observations supplemented by large-scale Potsdam Multi-Aperture Spectrophotometer/Pmas fiber PAcK spectroscopy in the other three.
All six measurements made use of near-IR Hubble Space Telescope images to characterize the stellar contribution to the gravitational potential in the targets, allowing measurements of stellar masses based on Multi-Gaussian Expansion fits.

Although these galaxies are located in the local Universe ($z\lesssim0.02$), they are passively evolved relics of $z\sim2$ red nugget galaxies, with evidence displayed in their morphology and stellar populations that they have not undergone any significant growth to their stars or BHs---including no intermediate or major mergers---since $z\sim2$ \citep{Yildirim2017}.
These systems thus likely remain representative of common quiescent red nugget galaxies at $z\sim2$.
One advantage of dynamical \mbh\ measurements in these inactive relic galaxies is that they avoid the systematics affecting single-epoch measurements by directly detecting the bulk motion of stars or molecular gas within the BH SOI.
Measurements in this sample also include thorough explorations of systematic uncertainties in \mbh, in additional to statistical uncertainties.
Here, we present their total measurement uncertainties, with statistical and systematic sources of error summed in quadrature.
These direct detections are the most robust \mbh\ measurements made in galaxies representative of populations outside the local Universe.

We also include the first and thus-far only stellar-dynamical measurement in a gravitationally lensed quiescent galaxy at $z=1.95$ \citep{Newman2025}.
This measurement used JWST/NIRSpec IFU, NIRCam imaging, and the Hubble Space Telescope/ACS imaging to resolve stellar kinematics in the galaxy disk and reconstruct the galaxy in the source plane.
They found the galaxy is massive and compact, reflecting the properties of typical red nuggets at cosmic noon.
The exceptional lensing magnification (a factor of 29) and 91 pc resolution allowed for the robust measurement of \mbh\ and stellar mass with Jeans anisotropic modeling \citep{Newman2025}.

At higher redshifts, we include multiple AGN-based \mbh\ samples, including a sample of luminous QSOs at $5.9 < z < 7.1$ \citep{Yue2024}, broad-line selected AGN at $4 < z < 11$ \citep{Maiolino2024}, and LRDs at $5<z<8$ \citep{Durodola2025}.
We choose these samples to represent a variety of galaxy and measurement types at high-redshift, given debate in the literature regarding the best approach for estimating high-$z$ BH masses.
In \citet{Yue2024}, the QSOs were confirmed with NIRSpec spectroscopy and \mbh\ was estimated using H$\alpha$ following \citet{Reines2013}.
\citet{Maiolino2024} presented GN-z11 and other AGN that are not luminous QSOs, identified in the JWST Advanced Deep Extragalactic Survey with NIRSpec.
They estimated \mbh\ from broad-line H$\alpha$ or H$\beta$, following \citet{Reines2013}.
\citet{Maiolino2024} also estimated the stellar velocity dispersion ($\sigma_\star$) by taking the ionized gas velocity dispersion to be a reliable proxy of $\sigma_\star$, with a small correction following \citet{Bezanson2018}, enabling their AGN to be compared to the \mbh$-$stellar velocity dispersion relation (\msig).
Finally, \citet{Durodola2025} examined LRDs with NIRCam and MIRI, using the spectral energy distribution (SED)-fitting code \texttt{CIGALE} to estimate the AGN contribution to the SED and reporting the corresponding best-fit \mbh.
Although their method lacks spectroscopic confirmation, they found MIRI coverage was vital for deciphering between AGN and non-AGN contributions to the SEDs of LRDs.

\section{\label{results}Results}
\begin{figure}
\centering
\includegraphics[width=0.47\textwidth]{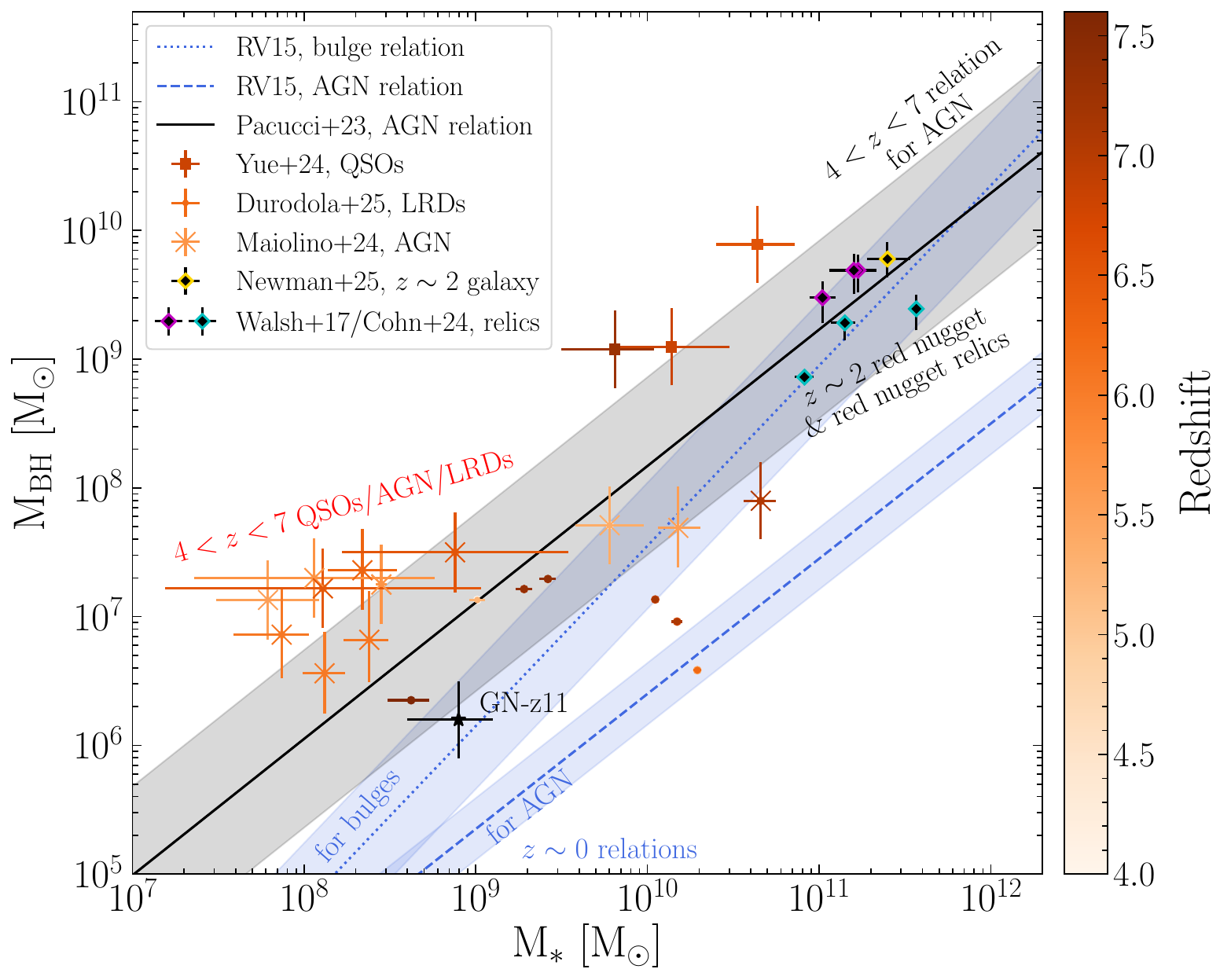}
\caption{The \mmtot\ relationship at high-$z$ and $z\sim0$.
AGN at $4\lesssim z\lesssim7$ are colored by their redshift as per the scale bar at right.
QSOs from \citet{Yue2024} are plotted as squares, AGN from \citet{Maiolino2024} are plotted as Xs, and LRDs from \citet{Durodola2025} are plotted as dots.
GN-z11, as measured in \citet{Maiolino2024nature}, is shown as a black star.
The stellar-dynamical \mbh\ measured in a $\sim2$ lensed quiescent galaxy is plotted as a black diamond outlined in yellow \citep{Newman2025}.
Dynamical \mbh\ measurements in red nugget relic galaxies---local galaxies representative of common quiescent galaxies at $z\sim2$---are plotted as black diamonds outlined in cyan and magenta for molecular gas-dynamical \citep{Cohn2021,Cohn2023,Cohn2024} and stellar-dynamical measurements \citep{Walsh2015,Walsh2016,Walsh2017}, respectively.
The \citet{ReinesVolonteri2015} $z\sim0$ \mmtot\ relations for AGN (blue dashed line) and bulges (blue dotted line) are shown alongside the \citet{Pacucci2023} $4<z<7$ \mmtot\ AGN relation (black solid line).
Shaded regions represent the intrinsic scatter of the relations.
The \citet{Yue2024} and \citet{Maiolino2024} samples follow the \citet{Pacucci2023} relation more closely than the $z\sim0$ relations, while the \citet{Durodola2025} LRDs are scattered across all three relations.
Remarkably, the red nugget relics and the $z\sim2$ quiescent galaxy are aligned with both the high-$z$ \citet{Pacucci2023} relation and the $z\sim0$ \citet{ReinesVolonteri2015} relation for bulges.
}
\label{fig_mm}
\end{figure}

First, we present the BH masses from our samples on the \mmtot\ relations (Figure \ref{fig_mm}).
To make the comparison between local and high-$z$ relations, we include the \citet{ReinesVolonteri2015} and \citet{Pacucci2023} relations.
\citet{ReinesVolonteri2015} use samples of broad-line AGN with single-epoch \mbh\ measurements following \citet{Reines2013}, dwarf galaxies with broad-line AGN from \citet{Reines2013}, reverberation mapped AGN from \citet{Bentz2015}, and dynamical \mbh\ measurements from \citet{Kormendy2013}.
They then build one local relation for AGN hosts and another local relation for the dynamically-measured ellipticals and classical bulges.
\citet{Pacucci2023} select AGN at $4<z<7$ from \citet{Harikane2023}, \citet{Maiolino2024}, and \citet{Ubler2023}, and calculate BH masses following \citet{Reines2013}.
They calculate a new \mmtot\ relation based on these data, as well as an estimate of the high-$z$ galaxy stellar mass function to account for detection bias.

The majority of the $4<z<11$ AGN, QSOs, and LRDs in our sample are consistent within the scatter of the \citet{Pacucci2023} relation.
Only one LRD lies within the scatter of the \citet{ReinesVolonteri2015} relation for AGN, with all other BHs in the sample significantly overmassive compared to that relation.
The \citet{Yue2024} QSOs are slightly overmassive even compared to the high-redshift relation, with the majority of the \citet{Maiolino2024} AGN also offset above the relation.

Strikingly, the $z\sim2$ quiescent galaxy and the red nugget relics are all consistent with both the \citet{Pacucci2023} high-$z$ relation and the $z\sim0$ \citet{ReinesVolonteri2015} \mmtot\ relation for bulge-dominated galaxies.
This is perhaps unsurprising, as red nuggets are likely bulge-dominated objects, although there has been some disagreement about their bulge decomposition (e.g., \citealt{Graham2016b,SavorgnanGraham2016}).
The relics were previously found to be consistent with the \citet{Zhu2021} \mbh$-M_\mathrm{core}$ relation \citep{Cohn2024}, reflecting the fact that $z\sim2$ red nuggets are thought to become the cores of local massive ETGs \citep{Dokkum2010}.
The red nuggets' agreement with the high-$z$ AGN relation could also suggest the $4<z<7$ \mmtot\ relation may remain in place at $z\sim2$.

\begin{figure}
\centering
\includegraphics[width=0.47\textwidth]{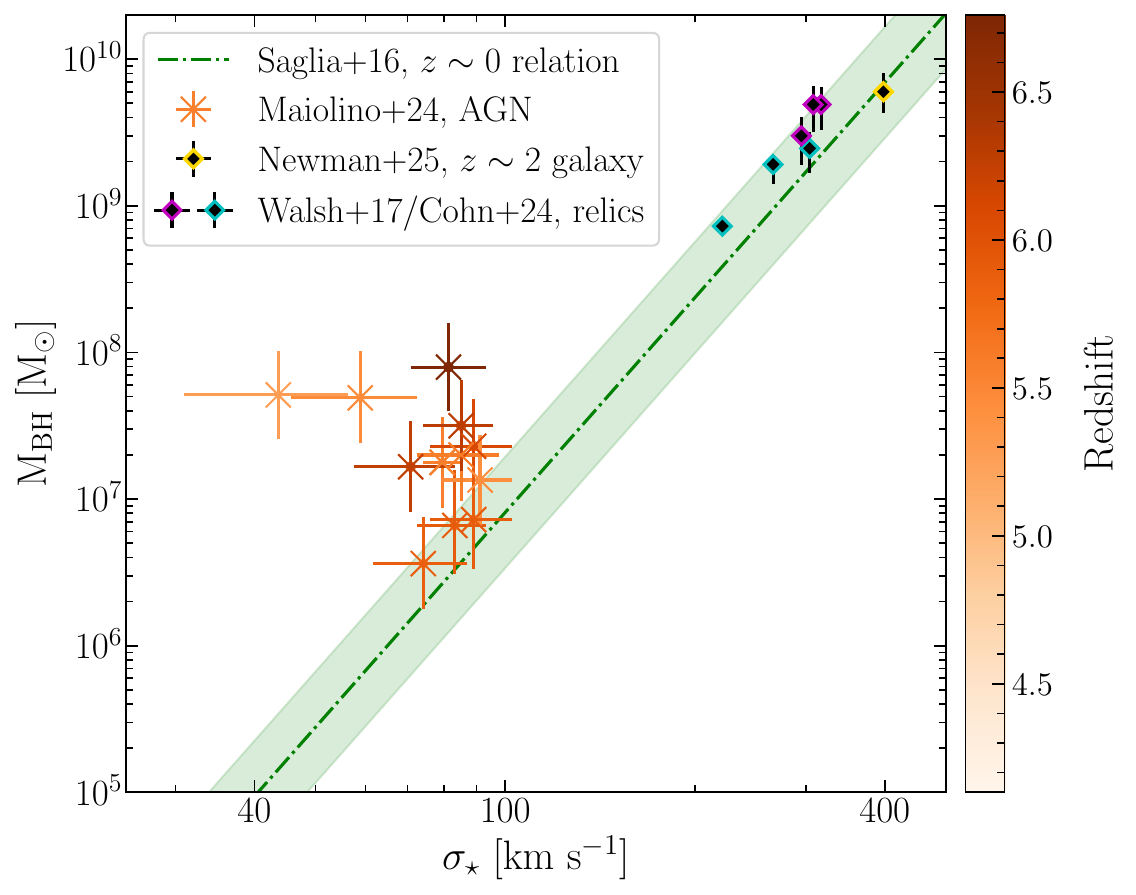}
\caption{High-redshift AGN and $z\sim2$-representative galaxies on the \msig\ relation.
Symbols are the same as in Figure \ref{fig_mm}.
\citet{Maiolino2024} is the only high-$z$ sample with estimated stellar velocity dispersions.
The $z\sim0$ \msig\ relation (dash-dot green line; \citealt{Saglia2016}) is also shown, with the shaded region showing its intrinsic scatter.
Although the AGN and red nugget relic galaxy populations overlap with the $z\sim0$ relation, every galaxy is offset above the median of the relation except for the lensed $z\sim2$ system from \citet{Newman2025}.
}
\label{fig_msig}
\end{figure}

In Figure \ref{fig_msig}, we compare the red nugget galaxies and the \citet{Maiolino2024} AGN to the \msig\ relation.
Although all of the red nugget galaxies are consistent with the relation within its scatter, every object in each sample is over-massive compared to the relation itself, except for the lensed quiescent galaxy at $z\sim2$.
This could point toward tentative evidence for a small positive evolution of the \msig\ relation.

We also perform a 10,000-iteration Monte Carlo to characterize how well the high-$z$ AGN, $z\sim2$ quiescent galaxy, and red nugget relic galaxies align with (or how much they differ from) the local and high-redshift \mmtot\ relations.
Each iteration, we sample the measured \mbh\ and $M_\star$ for each of our objects within their uncertainties.
Using the sampled $M_\star$ for each object, we then draw a random \mbh\ from a distribution centered on the \mmtot\ scaling relation with a width equal to the relation's intrinsic scatter.
Next, we calculate the ratio of the randomly sampled measurement \mbh\ to the randomly sampled scaling relation \mbh.
Finally, for each iteration, we calculate the median of that ratio for all objects in our sample.

From this Monte Carlo, we find that the median sampled \mbh\ is a factor of 0.7 times as massive as the \citet{Pacucci2023} prediction.
In all 10,000 iterations, the median sampled \mbh\ is consistent within the scatter of the $4<z<7$ \mmtot\ relation.
Conversely, we find that the median sampled \mbh\ measurement is 45 times more massive than the \citet{ReinesVolonteri2015} AGN-based relation prediction.
In all 10,000 iterations, the median sampled \mbh\ is overmassive beyond the scatter of that relation.
However, we also find that the median sampled \mbh\ is only 3 times more massive than the \citet{ReinesVolonteri2015} $z\sim0$ bulge-based relation prediction, and in 78\% of iterations, the median sampled \mbh\ is within the scatter of that relation.

Finally, we fit a new line to our combined samples of QSOs, AGN, LRDs, the lensed quiescent galaxy at $z\sim2$, and the red nugget relic galaxies, treating them as representative of a possible single $z\gtrsim2$ \mmtot\ relation.
Following both \citet{ReinesVolonteri2015} and \citet{Pacucci2023}, we assume the relation follows the form of $\log_{10}$\mbh$=m\log_{10}M_\star + b$.
We find $m=1.22\pm0.20$ with $b=-3.82\pm1.84$.
This line is slightly steeper than the \citet{Pacucci2023} relation ($m=1.06\pm0.09$) and slightly shallower than the \citet{ReinesVolonteri2015} bulge relation ($m=1.40\pm0.21$), but within the high-$z$ AGN relation's intrinsic scatter.
However, as we discuss in \S\ref{conclusions}, we caution that our results may point toward evidence for differences in scaling relations due to diverging evolutionary histories, rather than a single $2<z<7$ \mmtot\ relation.

\begin{figure*}
\centering
\includegraphics[width=\textwidth]{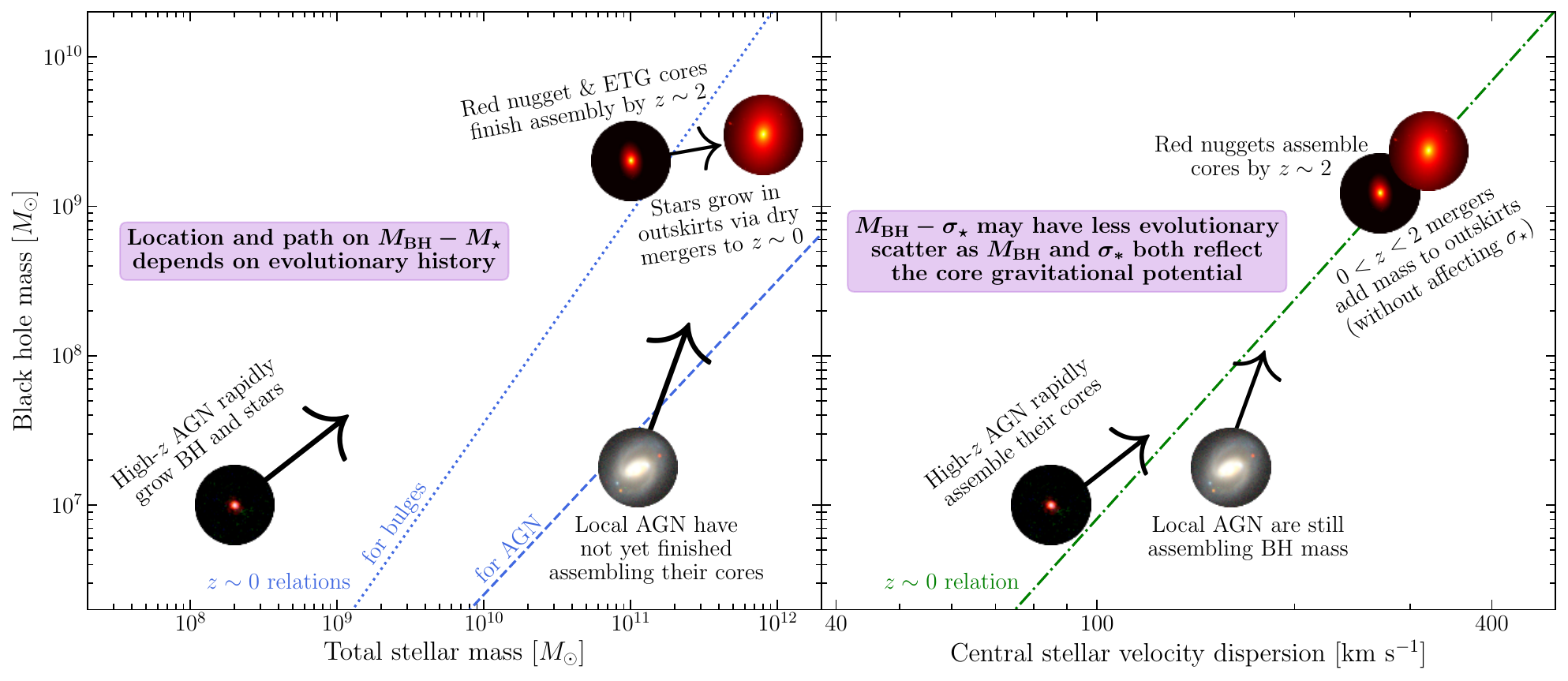}
\caption{Schematic showing how different evolutionary pathways may determine the locations of galaxies in the \mmtot\ (left) and \msig\ (right) parameter spaces.
On \mmtot: red nugget relics and compact $z\sim2$ quiescent galaxies exist on the $z\sim0$ relation for bulges, reflecting that \mbh\ growth may correspond with the assembly of galaxy bulges.
However, total stellar mass includes stars assembled beyond the bulge, indicating different assembly histories may lead to different positions on the \mmtot\ diagram.
AGN systems at high-$z$ are accreting material, rapidly growing their BHs and forming new stars in a regime where much of the galaxy mass likely remains in gas reservoirs.
Red nuggets assemble their cores by $z\sim2$, then grow into local ETGs through dry mergers that increase total stellar mass without feeding the bulge or BH.
AGN in the nearby Universe have undergone a different growth history, assembling large total stellar masses without completing their core or BH growth.
Included in the panel for guidance are the $z\sim0$ relations from \citet{ReinesVolonteri2015}.
On \msig: both \mbh\ and $\sigma_\star$ reflect the central gravitational potential of the galaxy, explaining why this relation may change less with redshift or diverging evolutionary histories.
Shown in this panel is the \citet{Saglia2016} \msig\ relation.
}
\label{fig_schematic}
\end{figure*}

\section{Discussion and Conclusions\label{conclusions}}
There has been substantial debate regarding overmassive BHs in the early Universe.
Studying relic galaxies and gravitationally lensed objects present the only means of exploring this regime through direct \mbh\ detection, without relying solely on AGN-derived BH masses.
We find that local massive red nugget relic galaxies, likely representative of the typical quiescent galaxy population at $z\sim2$, are consistent with the local \citet{ReinesVolonteri2015} \mmtot\ relation for bulges.
The $z\sim2$ lensed quiescent galaxy also aligns with this relation.
Given that red nugget galaxies are likely bulges themselves, this is perhaps unsurprising and suggests that the \mmtot\ relation for bulges may have been in place as early as $z\sim2$.
Our results thus indicate that differences in the BH scaling relations are likely driven by differing evolutionary pathways.

In this case, the position of different galaxy populations in the \mmtot\ parameter space would vary significantly due to the myriad evolutionary processes represented.
The processes that drive growth in the cores and bulges of galaxies differ from those that drive growth in the outskirts (e.g., \citealt{Dokkum2010}), meaning BHs (which likely grow alongside cores and bulges) will not coevolve consistently with total $M_\star$ (which is also affected by processes outside cores and bulges).
The recent findings that AGN host overmassive BHs at high redshift could thus be reflective of different dominant evolutionary processes in the host galaxies.
This is illustrated in the left panel of Figure \ref{fig_schematic}, which shows galaxies throughout the Universe's history undergoing different regimes of growth and occupying different regions of \mmtot\ parameter space.

In support of this picture, BHs have been found to correlate more strongly with different host properties (e.g., $M_\star$, $M_\mathrm{bul}$, or $\sigma_\star$) depending on galaxy classification \citep{Kormendy2013,Saglia2016,Graham2025}.
Interestingly, recent simulations also suggest scatter in the \mmtot\ relation may be reduced as a result of hierarchical merging of quenched systems, a process that may not regularly occur until $z<2$ \citep{Zhu2025}.
Of course, if variations in evolutionary history drive differences in the placement of systems on \mmtot, the overall relation will likely evolve with redshift due to different stages of growth dominating at different times in the Universe.
It is also worth noting that the red nugget relics and $z\sim2$ quiescent galaxy remain overmassive compared to the \citet{Kormendy2013} $z\sim0$ \mmass\ relation, but \citet{ReinesVolonteri2015} and \citet{Kormendy2013} make different assumptions about the stellar initial mass function, resulting in the relations' offset.

Intriguingly, the $z\sim2$ quiescent galaxy and red nugget relic galaxies also lie on the \citet{Pacucci2023} \mmtot\ relation for AGN at $4<z<7$.
We fit a line to a sample consisting of these $z\sim2$-representative galaxies and the $4<z<7$ QSOs, LRDs, and AGN, finding it agrees with the \citet{Pacucci2023} relation within uncertainties.
This could provide evidence that this $4<z<7$ \mmtot\ relation was in place from $2<z<7$.
If this is the case, typical evolution processes between $0<z<2$ (e.g., hierarchical merging) would be required to provide more stellar growth than BH growth to produce the local AGN relation.
The agreement of the local red nugget relics and the $z\sim2$ quiescent galaxy with the $4<z<7$ AGN relation could also suggest that AGN with swiftly growing BHs in the early Universe may be progenitors of $z\sim2$ red nuggets.

However, we caution that there is no definitive proof the red nuggets and high-$z$ AGN follow a single evolutionary pathway.
As such, we do not place our own line on Figure \ref{fig_mm}, instead simply presenting it as a test case that shows consistency with \citet{Pacucci2023}.
In general, the consistency of the relic galaxies and the lensed $z\sim2$ galaxy with both the high-$z$ AGN relation and the $z\sim0$ bulge relation demonstrates the importance of carefully considering different evolutionary pathways when comparing galaxies to the many existing relations.

Finally, our comparison of red nugget galaxies and high-$z$ AGN to the \citet{Saglia2016} $z\sim0$ \msig\ relation could provide evidence for a slight positive offset from this relation at high redshifts, although all six relic galaxies and the lensed $z\sim2$ galaxy remain consistent with the relation within its scatter.
As demonstrated in the right panel of Figure \ref{fig_schematic}, since \mbh\ and $\sigma_\star$ reflect the gravitational potential at the center of the galaxy, the \msig\ relation may be more fundamental regardless of changing evolutionary stages or pathways.
However, a large region of parameter space in velocity dispersion ($100$ \kms$\lesssim\sigma_\star\lesssim200$ \kms) remains unexplored beyond the local Universe.
Obtaining more $\sigma_\star$ estimates for AGN, identifying more relic galaxies, and making \mbh\ measurements in more gravitationally lensed galaxies will serve to fill in the gaps and make a more confident comparison to the local \msig\ relation.

As we demonstrate here, dynamical \mbh\ measurements in relic galaxies are an important tool for probing the scaling relations in the early Universe via a complementary, independent method from AGN.
In the future, explorations into the nature of LRDs and AGN at high-$z$ will be critical for improving our understanding of BH$-$host galaxy coevolution.
Identifying and making more dynamical \mbh\ measurements in red nugget relic galaxies, as well as measuring more dynamical BH masses in $z\gtrsim2$ lensed galaxies, will further clarify the state of the scaling relations at $z\gtrsim2$, providing an important stepping stone between local and high-$z$ BH scaling relations.

\section*{Acknowledgments}
This work used arXiv.org and NASA's Astrophysics Data System for bibliographic information.
R.C.H. acknowledges support from NASA under ADAP grant 80NSSC23K0485. % TO DO: add Ryan's next acknowledgment soon!
J.H.C. would like to thank Jonathan Trump (no relation) for the encouraging discussion that led to this paper, Taylor Hutchison for helpful advice on figure design, and Addy Evans and Silvana Delgado Andrade for moral support.

\software{MATPLOTLIB \citep{Hunter2007}, NUMPY \citep{Walt2011,Harris2020}, SCIPY \citep{Virtanen2020}.}

\end{document}